  \documentstyle[twocolumn,prl,aps,epsfig,epsf,amssymb,amsfonts,psfrag]{revtex}
\begin{document}
%
%
\renewcommand{\Re}{\operatorname{Re}}
\renewcommand{\Im}{\operatorname{Im}}
\newcommand{\Tr}{\operatorname{Tr}}
\newcommand{\sign}{\operatorname{sign}}
\newcommand{\dd}{\text{d}}
\newcommand{\q}{\boldsymbol q}
\newcommand{\p}{\boldsymbol p}
\newcommand{\rr}{\boldsymbol r}
\newcommand{\pp}{p_v}
\newcommand{\vv}{\boldsymbol v}
\newcommand{\I}{{\rm i}}
\newcommand{\pphi}{\boldsymbol \phi}
\newcommand{\ds}{\displaystyle}
\newcommand{\be}{\begin{equation}}
\newcommand{\ee}{\end{equation}}
\newcommand{\bea}{\begin{eqnarray}}
\newcommand{\eea}{\end{eqnarray}}
\newcommand{\Acl}{{\cal A}}
\newcommand{\Rcl}{{\cal R}}
\newcommand{\Tcl}{{\cal T}}
\newcommand{\Tmin}{{T_{\rm min}}}
\newcommand{\Toff}{{\langle \delta T \rangle_{\rm off} }}
\newcommand{\Roff}{{\langle \delta R \rangle_{\rm off} }}
\newcommand{\RoffI}{{\langle \delta R_I \rangle_{\rm off} }}
\newcommand{\RoffII}{{\langle \delta R_{II} \rangle_{\rm off} }}
\newcommand{\dg}{{\langle \delta g \rangle_{\rm off} }}
\newcommand{\rd}{{\rm d}}
\newcommand{\br}{{\bf r}}
\newcommand{\la}{\langle}
\newcommand{\ra}{\rangle}

\twocolumn[\hsize\textwidth\columnwidth\hsize\csname @twocolumnfalse\endcsname
%
%
\draft

\title{Conditions for Adiabatic Spin Transport in Disordered Systems}

\author{Markus Popp$^{(1)}$, Diego Frustaglia$^{(2)}$,
        and Klaus Richter$^{(1)}$ }

\address{
$^{(1)}$ Institut f{\"u}r Theoretische Physik, Universit{\"a}t Regensburg, 
         93040 Regensburg, Germany}

\address{
$^{(2)}$ Institut f{\"u}r Theoretische Festk{\"o}rperphysik,
 Universit{\"a}t Karlsruhe, 76128 Karlsruhe, Germany }

\date{\today}

\maketitle
\begin{abstract}
We address the controversy concerning the necessary conditions for the observation 
of Berry phases in disordered mesoscopic conductors. For this purpose we 
calculate the spin-dependent conductance of disordered two-dimensional
structures in the presence of inhomogeneous magnetic fields. Our numerical 
results show that for both, the overall conductance and quantum corrections,
the relevant parameter defining adiabatic spin transport scales with 
the square root of the number of scattering 
events, in generalization of Stern's original proposal\cite{S92}.
This could hinder a clear-cut experimental observation 
of Berry phase effects in diffusive metallic rings.
\end{abstract}


]

\narrowtext

In contrast to phenomena related to Aharonov-Bohm (AB) phases \cite{AB59} 
for charge carriers, the corresponding observation of Berry phases 
\cite{B84} due to the 
coupling of a spin to an orientationally nonuniform magnetic field $B$
requires the limit of {\it adiabatic} spin evolution. In mesoscopic conductors,
such a limit corresponds to the situation where the carrier spin can follow 
the spatially varying field during transport through the system. 
In terms of time scales, the adiabatic limit is reached when the Larmor
frequency of spin 
precession, $\omega_{\rm s}=2 \mu B / \hbar$, is large compared to the 
reciprocal of a characteristic time scale $t_{\rm c}$ on which, from the point of 
view of the spin, the direction of the field has changed significantly during 
motion. There is consensus that for \emph{ballistic} (disorder-free)
systems with magnetic field configurations commonly theoretically considered 
\cite{S92,LSG93,FR01,FHR01} and experimentally realized\cite{YTW98},
it holds 
$t_{\rm c} \sim L/v_{\rm F}$, where $v_{\rm F}$ denotes the Fermi velocity of the 
carriers, and $L$ is the characteristic length scale of the system over 
which the field changes. For one-dimensional (1d) ballistic systems 
the condition  for adiabaticity,  $\omega_{\rm s} \gg 2\pi/ t_{\rm c}$,
therefore reads \cite{commentQ}
\begin{equation}
\label{Q}
Q_{\rm 1d} \equiv 
\frac{\omega_{\rm s}}{2 \pi v_{\rm F}/L} \gg 1 \; ,
\end{equation}
where we introduced the adiabaticity parameter $Q_{\rm 1d}$. 

However, in the case of {\em disordered} systems there are two candidates for 
the characteristic time $t_{\rm c}$: (i) the mean elastic scattering time $\tau$ 
and (ii) the Thouless time $t_{\rm Th}=(L/ \ell)^2 \tau$, with $\ell=v_{\rm F} \tau$ 
the elastic mean free path. The issue which of these two time scales is the 
relevant one has recently led to a controversial discussion \cite{LKPB99,LSG99}.

In his proposal for 
1d diffusive rings Stern \cite{S92} perturbatively calculated the lifetime of the adiabatic 
eigenstates and compared it to $t_{\rm Th}$.
He arrived at the condition 
\begin{equation}\label{Qstern}
Q_{\rm 1d}   \gg L/\ell
\end{equation}
for adiabatic spin transport (Eq.~(7) in Ref.\ \cite{S92}).
This corresponds to setting $t_c=\tau$. 
Comparing Eqs.\  (\ref{Q}) and (\ref{Qstern})  one
recognizes that in the diffusive regime, $L\gg \ell$, the adiabatic limit
would require a magnetic field $L/ \ell$ times larger than in the ballistic 
case. This  ``pessimistic criterion'', which later has also been advocated by 
van Langen \emph{et al.\ } \cite{LKPB99}, 
would imply field strengths in the quantum 
Hall regime that let an experimental observation of 
Berry phases in diffusive metallic rings appear rather unlikely.

Alternatively, in analogy to the ballistic travelling time $L/v_F$,
it appears convincing to associate $t_{\rm c}$ for diffusive systems 
with $t_{\rm Th}$, the time it takes the electron to 
\emph{diffuse} through the structure. 
This argumentation has been put forward
by Loss \emph{et al.\ } \cite{LSG93,LSG99,EL00}. By calculating  
the quantum corrections of the  conductance in diffusive 1d rings 
they predicted clear signatures of Berry's phase to be observable 
in a regime given by
\begin{equation}\label{Qloss}
Q_{\rm 1d} \gg \ell/L \, .
\end{equation}
This condition for adiabaticity differs from criterion
(\ref{Qstern}) 
by a factor $(L/ \ell)^2$ and predicts for the observability of Berry phases
a field strength above 20 mT\cite{EL00}, which is well in 
reach of modern experimental techniques \cite{YTW98}.

In view of various recent experimental efforts to observe Berry phases
in the magneto conductance of mesoscopic rings \cite{YTW98,PhDH,Wees,Yau},
a clarification of this issue of the relevant time scale is desirable.
The derivations of the conditions (\ref{Qstern}) and (\ref{Qloss}) were based 
on diagrammatic and semiclassical techniques. Here we choose a 
different approach and study numerically the spin-dependent  conductance
of ballistic and disordered mesoscopic systems in the presence of a 
spatially 
varying magnetic field $\vec B(\vec r)=\nabla \times \vec A(\vec r)$. 
The Hamiltonian for noninteracting electrons with effective mass $m^*$ and 
charge $-e$ reads
\begin{equation}\label{H}
H=\frac{1}{2m^*}\left[\vec p + \frac{e}{c} 
\vec A(\vec r) \right]^2 + V(\vec r) + \mu\vec B(\vec r) \cdot\vec \sigma. 
\end{equation}
The nontrivial coupling of the spin to the magnetic field enters via the 
Zeeman term $\mu\vec B(\vec r) \cdot\vec \sigma$, where $\vec \sigma$ is the 
Pauli spin vector and $\mu = g^*e\hbar/(4m_0 c)$ the magnetic moment with 
$g^\ast$ the gyromagnetic ratio. 
The electrostatic potential $V(\vec r)$ includes the confinement  
and the potential of random impurities in the disordered case. 
At $T=0$ the spin-dependent conductance of a mesoscopic system with two 
attached leads is given by the Landauer formula \cite{FG97}
\begin{equation} \label{LF}
G =\frac{e^2}{h} \sum_{s',s=\pm 1} T_{s's} = \frac{e^2}{h}\; 
\sum_{m', m=1}^M  \sum_{s',s=\pm 1} |t^{m' m}_{s's}|^2  \; .
\end{equation}
Here $t^{m' m}_{s's}$ is the transmission amplitude 
from an incoming channel $m$ with spin $s$ to an outgoing channel ($m', s'$).
We calculate $t^{m' m}_{s's}$ by projecting the 
corres\-ponding Green function matrix onto the asymptotic spi\-nors in the leads. 
We compute the Green function for the Hamiltonian (\ref{H}) numerically, 
using a generalized version of the recursive Green function technique based 
on a tight-binding model \cite{FG97} including spin \cite{commentGFT}. 
We model the (non-magnetic) disorder potential leading to elastic scattering
within an Anderson model by chosing random delta-like scatterers 
with amplitudes following a box distribution. The spin-dependent conductance 
is then obtained from ensemble averages over independent disorder configurations
\cite{comment-method}.

\begin{figure}[h]
        \psfrag{S}{{\small $\mathrm{\vec S_{in}}$}}
        \psfrag{Sf}{{\small $\mathrm{\vec S_{out}}$}}
        \psfrag{B}{{\small $\mathrm{\vec{B}}$}}
        \psfrag{L}{{\small $\mathrm{L}$}}
        \psfrag{W}{{\small $\mathrm{W}$}}
        \psfrag{v}{{\small $\mathrm{\vec v}$}}
        \begin{center}
      \epsfig{file=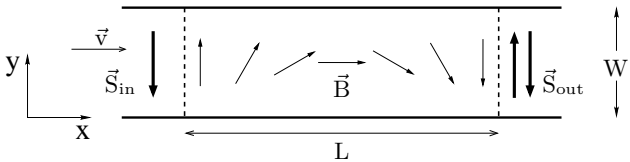,width=0.95\linewidth}       
       \end{center}
\caption{2d strip configuration used for the 
calculations of the spin-dependent conductance. The magnetic field 
$\vec{B}(x)$ performs a $180^\circ$-rotation within the plane of the 
strip. Spin states ($\vec{S}_{\rm in}$, $\vec{S}_{\rm out}$)
are defined with respect to the $y$-axis.}
       \label{strip}
\end{figure}

We now turn to the subject of interest and study how adiabaticity is 
approached in mesoscopic spin quantum transport. For this purpose we introduce a 
model system consisting of a 2d strip with a rotating in-plane magnetic field 
between two ballistic leads, see Fig.~\ref{strip}. This sys\-tem can also
be regarded as a model for transport through magnetic domain walls.
We assume incoming 
electrons with spin-down polarization in the $-y$ direction\cite{comment-spinup},
 in\-jec\-ted from the left 
 with Fermi wave number $k_{\rm F} = 2\pi/\lambda_{\rm F}$.

\begin{figure} 
\begin{center}
  \psfrag{Q}{{ $\mathrm{Q}$}}
  \psfrag{T}{{ $\mathrm{\langle T_{\downarrow \downarrow}\rangle }$}}
\psfig{figure=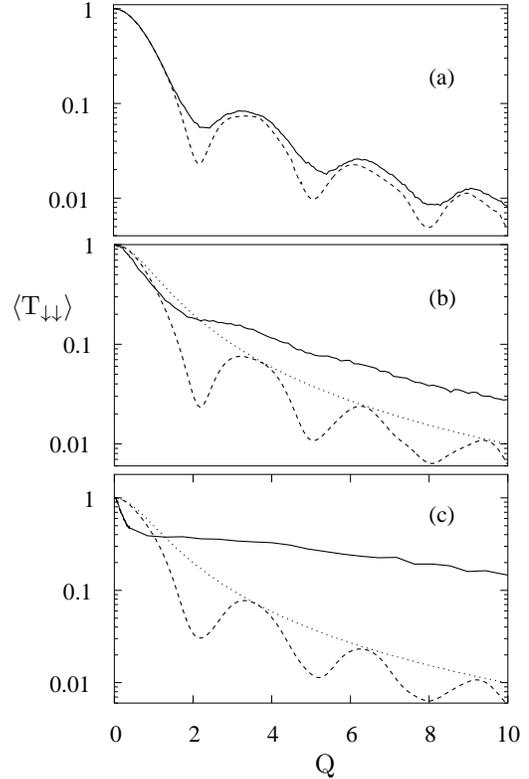,width=0.8\linewidth,angle=0} 
 \end{center}
\caption{
Ensemble averaged normalized transmission 
$\langle T_{\downarrow \downarrow}\rangle$
for a  disordered (solid lines) and 
 ballistic (dashed)  2d strip (Fig.~\ref{strip}) 
as a function of the adiabaticity parameter $Q\sim B$. 
The panels correspond to different disorder 
strengths: (a) quasi-ballistic: $L/ \ell =0.5$ ($L/W=4.4$), 
           (b) moderate: $L/ \ell =3$ ($L/W=7.8$), 
	   and (c) diffusive: $L/ \ell =10$ ($L/W=13.9$). 
The ballistic curves, Eq.~(\ref{T2d}), differ slightly from each 
other since they correspond to different Fermi wave vectors
(ranging from $k_{\rm F} W/\pi = 7.7$ to 11.6) but 
show the same overall Lorentzian decay with $Q$ (dotted).
}
\label{compare}
\end{figure}

We first consider the overall conductance.
In the ballistic case, it is feasible to derive an analytical expression 
for the spin-resolved transmission of this system using a transfer matrix 
approach. 
The normalized transmission $T_{\downarrow \downarrow}$ for spin-down
polarized incoming electrons to exit the system with spin-down polarization
reads 
\cite{markus}
\begin{equation}\label{T2d}
  T_{\downarrow  \downarrow }\equiv \frac{1}{M} \! \sum_{m',m=1}^M  \!
  |t^{m' m}_{-1, -1}|^2 \nonumber 
  = \sum_{m=1}^M \ \frac{\sin^2\left(\frac{\pi}{2}
   \sqrt{1 +Q_m^2} \right )}{M( 1+Q_{m}^2) } 
\end{equation}
with the generalized adiabaticity parameter (Eq.~(\ref{Q})),
\be
 Q_m \equiv 
 \frac{g^*}{k_{\rm F}W}  
  \frac{m^*}{m_0} 
 \left(\frac{LWB}{hc/e} \right) \
 \left[1- \left(\frac{ m \pi}{k_{\rm F} W}\right)^2 \right]^{-1/2}   \; ,
\ee
for the $m$-th propagating mode in a 2d strip of length $L$ and width $W$.  
Summing over  all transverse modes in Eq.~(\ref{T2d}) we find that 
the overall dependence of the ballistic transmission (dashed lines
in Fig.~\ref{compare}) is given by a Lorentzian
$ T_{\downarrow  \downarrow }\simeq1/(1+Q^2)$ (dotted lines). This defines  
an `effective' adiabaticity parameter $Q\sim B$ for the 2d strip, with $Q_1 < Q < Q_M$
and $Q \sim 1.4 \ Q_{\rm 1d}$. 
This allows us 
to introduce a quantity that solely characterizes the adiabatic regime 
in the case of several open  channels. 

For $B \! \to\! 0$ the spin direction is preserved and $ T_{\downarrow
\downarrow}$ is maximal.
In the limit of a strong $B$-field the spin stays adiabatically
aligned with the orientationally inhomogeneous field during transport, 
minimizing the probability of leaving the conductor in Fig.~\ref{strip} 
in a spin-down state. The Lorentzian dependence of 
$T_{\downarrow  \downarrow }$ on $Q\sim B$ reflects this behavior 
and appears as the natural measure for the crossover from the non-adiabatic 
($T_{\downarrow  \downarrow } \to 1, Q\ll 1$) to the adiabatic 
($T_{\downarrow  \downarrow } \to 0, Q\gg  1 $) regime. 

To find a proper condition for adiabaticity in the disordered case we 
 compute the ensemble averaged transmission $\langle T_{\downarrow  
\downarrow }\rangle$ in the presence of elastic scattering for $\lambda_{\rm F} \ll 
\ell$ as a function of $Q$ and compare it to the ballistic result~(\ref{T2d}).
Our results for different ratios $L/ \ell$ are shown as the solid lines in 
Fig.~\ref{compare} exhibiting the following features:\\
(i) The oscillations in the ballistic transmission are averaged out 
with increasing disorder.
(ii) For $Q\gg 1$, the normalized $\langle T_{\downarrow \downarrow}\rangle$ is 
larger in the disordered than in the ballistic case. This means that in 
the presence of elastic scattering a stronger scaled field $Q$ is required 
for acceding to the adiabatic regime of $\langle T_{\downarrow 
\downarrow}\rangle \approx 0 $.
(iii) For $Q\ll  1$ we observe the opposite behavior:
The non-adiabatic limit of almost maximum transmission $\langle 
T_{\downarrow \downarrow}\rangle $ is restricted to lower magnetic fields 
compared to the ballistic case.
(iv) The crossover region $Q\sim 1$ is characterized by a transmission plateau, 
which approaches  
$\langle T_{\downarrow \downarrow}\rangle \approx 
\langle T_{\uparrow \downarrow}\rangle \approx 0.5 $ with increasing diffusiveness. 
Here, the  non-magnetic 
disorder acts as a spin randomizer of the originally spin-polarized current. 
The features (i)-(iv) already begin to appear in the 
quasi-ballistic regime (Fig.~\ref{compare}(a)) and become more pronounced 
with increasing degree of diffusiveness given by the ratio $L/ \ell$ 
(Fig.~\ref{compare}(c)). 

\begin{figure} [h]
\begin{center}
\psfrag{N}{{ $ \langle \mathrm{N}\rangle $}}
\psfrag{L}{{ $ \mathrm{L/ \ell }$}}
\psfrag{Tad}{{\tiny $\langle T_{\downarrow \downarrow}\rangle$}}
\psfrag{$T_{\downarrow \downarrow}$}{{\tiny $\langle T_{\downarrow \downarrow}\rangle$}}
\epsfig{figure=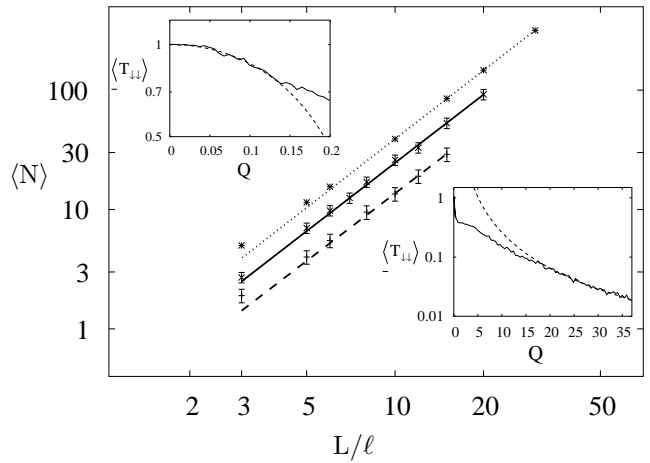,width=0.7\linewidth,angle=-90} 
 \end{center}
\caption{
Functional dependence of fit parameters $N_{\rm ad}$ 
(solid line) and $N_{\rm nad}$ (dashed) on the scaled length $L/ \ell$ of 
the disordered strip. Linear regression yields 
$N_{\rm ad} = 0.31 \ (L / \ell )^{1.9}$ (solid line) and
$N_{\rm nad} = 0.18 \ (L / \ell )^{1.88}$ (dashed). 
For comparison the mean number of scattering events 
$\langle N\rangle=0.48 \ (L / \ell )^{1.91} $ is also shown 
(dotted) obtained from an independent 1d random walk model.
Insets: Transmission 
$\langle T_{\downarrow \downarrow}\rangle$ and fitted Lorentzians in the 
nonadiabatic (left) and adiabatic (right) limit 
for strip with $L/ \ell=10$.
(Error bars include uncertainty from fitting procedure.) 
}
\label{pl}
\end{figure}

After this qualitative discussion we now derive a quantitative condition 
for adiabaticity in the disordered strip.
In analogy to the ballistic case we expect the disorder averaged transmission 
$\langle T_{\downarrow \downarrow}\rangle $ to exhibit a scaled Lorentzian
 dependence in the limits of small and large $Q$. Indeed,  in the adiabatic limit
the Lorentz function is an excellent fit to the diffusive 
curve $\langle T_{\downarrow \downarrow}\rangle $ in Fig.~\ref{compare}(c), 
if the ballistic parameter $Q$ 
is replaced by $Q/ \sqrt{N_{\rm ad}}$ with 
 $N_{\rm ad}$ being fitted (left inset Fig.\ \ref{pl}). 
 Corresponding results
hold for the nonadiabatic limit where we use the scaling
$Q \sqrt{N_{\rm nad}}$ (right inset in Fig.\ \ref{pl}).
We further determined  $N_{\rm ad}$ and $N_{\rm nad}$  
for various ratios $L/ \ell$ and obtain power-law dependences illustrated in 
Fig.~\ref{pl}. We  hence can formulate 
as a necessary condition for adiabatic spin transport through the 
disordered 2d strip: $Q\gg (L/ \ell)^{0.95}$. 
Comparing this with  Eq.~(\ref{Qstern}) we obtain a little smaller exponent. 
To explain this deviation we note that Eq.~(\ref{Qstern}) can  be 
written in the more general form $Q\gg \sqrt{\langle N\rangle}$, with
the average number of scattering events $\langle N\rangle=\langle
t_{\rm Th}\rangle/ \tau = (L/\ell)^2$. This suggests to associate
 $N_{\rm ad}$ and $N_{\rm nad}$ with the number of scattering events the 
electron has to undergo upon traversing the microstructure. Due to the strong 
coupling of the finite-size 2d strip  to the ballistic leads we expect 
 the diffusion time to be smaller than the Thouless time $t_{\rm Th}$, thus reducing
the number of scattering events.

To confirm the above arguments we independently checked numerically 
the dependence of 
$\langle N\rangle$ on the scaled length $L/ \ell$ of a finite disordered
conducting strip.
To this end we used
a 1d \cite{comment1d} random walk model 
taking explicitely into account the interface between 
disordered and ballistic regions. 
We find that $\langle N\rangle$ obeys a power law in
$L / \ell$ with exponent $1.91$ which, as expected, is lower than two
(dotted line in  Fig.~\ref{pl}).
In Fig.~\ref{pl}, besides small deviations for small $L / \ell$
in the non-diffusive limit,
there is good agreement with the fitted straight lines for
$L / \ell\gtrsim 5$, indicating diffusive behavior.
Within the given error tolerance 
all three curves in  Fig.~\ref{pl} exhibit identical exponents 
and deviate only in the prefactor of order one. 

We conclude from our numerical, quantum mechanical results,
together with the expression for $\langle N\rangle$
from the independent random-walk model, that the adiabaticity 
parameter scales with $\sqrt{\langle N \rangle}$.
This enables us to formulate a \emph{general} (system-independent) adiabaticity
 condition for  diffusive systems, that only depends on the corresponding
 adiabaticity parameter $Q$ of the ballistic system and the mean number of 
scattering events $\langle N\rangle$ \cite{commentmodel}: 
\begin{equation} \label{AC}
 Q\gg \sqrt{\langle N\rangle}\ .
\end{equation}
For diffusive 1d rings this criterion is in perfect agreement with Stern's original
condition (\ref{Qstern}). 
\vspace{-2mm}

\begin{figure} [h]
\begin{center}
\epsfig{figure=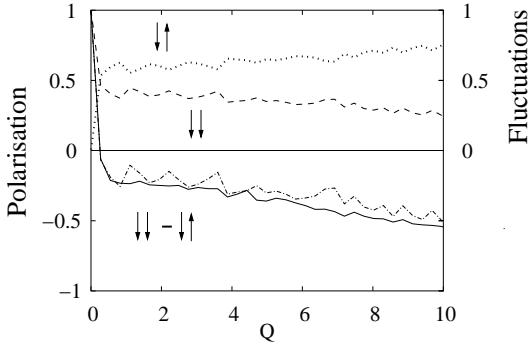,width=0.8\linewidth,angle=0} 
 \end{center}
\caption{
 Quantum fluctuations 
$\delta g_{\downarrow \downarrow}/(\delta g_{\downarrow \downarrow} 
+\delta g_{\uparrow \downarrow})$ 
(dashed) and $\delta g_{\uparrow \downarrow}/(\delta g_{\downarrow \downarrow} 
+\delta g_{\uparrow \downarrow})$ (dotted) as a
function of $Q$ for a diffusive strip with $L/ \ell =15$. For comparison we also
show $(\delta g_{\downarrow \downarrow} -\delta g_{\uparrow \downarrow}) / 
(\delta g_{\downarrow \downarrow} +\delta g_{\uparrow \downarrow})$ 
(dashed-dotted) and the polarization
 $(\langle T_{\downarrow \downarrow}\rangle  - 
 \langle T_{\uparrow \downarrow}\rangle )/(\langle 
 T_{\downarrow \downarrow}\rangle  + \langle T_{\uparrow \downarrow}\rangle ) $(solid).}
\label{fluc}
\end{figure}

So far we considered  the total
conductance which is dominated by the Boltzmann contribution.
However, signatures of Berry phases in diffusive conductors appear only
in the phase coherent part of the conductance, i.e. quantum corrections
such as Aharonov-Bohm (AB) oscillations and universal conductance fluctuations
(UCFs).  To decide whether distinct Berry phase effects, e.g.\ in diffusive rings,
can be observed at realistic magnetic field strengths, 
one has to check if an adiabaticity  condition different
from (\ref{AC}) holds for the quantum corrections.
For this purpose, spin-resolved UCFs represent a suitable quantity, defined
as $\delta g_{s's} = \sqrt{\langle T_{s's}^2\rangle -  \langle T_{s's}\rangle^2}$
in units of $e^2/h$.
We calculated $\delta g_{s's}$ numerically  as a function of $Q$
for a diffusive 2d strip with $L/ \ell =15$.
The results  are depicted in Fig.\ \ref{fluc} in terms of the normalized
difference $(\delta g_{\downarrow \downarrow} -
\delta g_{\uparrow \downarrow}) / (\delta g_{\downarrow \downarrow} +
\delta g_{\uparrow \downarrow})$ which can be regarded as a polarization.
We note that the UCFs exhibit precisely the same scaling behavior as
the corresponding
quantity for the total conductance, $(T_{\downarrow \downarrow} -
T_{\uparrow \downarrow})/
(T_{\downarrow \downarrow} + T_{\uparrow \downarrow})$, and consequently obey
condition (\ref{AC}). Fig. \ref{fluc} also
illustrates the important fact that in a wide region $\ell/L \lesssim Q
\lesssim L/ \ell $ the respective adiabatic ($T_{\uparrow \downarrow}$,
$\delta g_{\uparrow \downarrow}$) and  nonadiabatic ($T_{\downarrow \downarrow}$, $\delta
g_{\downarrow \downarrow}$) components are comparable in magnitude.

We further note that numerical quantum calculations of the spin-dependent
 magneto-conductance in  disordered  rings subject to a circular
inhomogeneous $B$-field show that signatures of Berry phases appear only in the 
AB-oscillations of the adiabatic  components ($  \langle T_{\uparrow \downarrow}\rangle $ and
$\delta g_{\uparrow \downarrow}$), which are dominated by electrons with
spin always aligned with the local field \cite{PFRunp}. In view of Fig.~\ref{fluc} we
hence can conclude for diffusive rings that in the experimentally relevant plateau
region $\ell /L 
\lesssim Q <  1$ the adiabatic components, which show Berry phase signatures 
are of the same   magnitude as the nonadiabatic components. Berry phase effects in
the AB-oscillations of the total conductance  and the UCF's are hence masked by 
the regular, nonadiabatic contribution 
\cite{comment-ring}. However, in this broad plateau region one still 
finds effects of the inhomogeneous magnetic field that can be ascribed to the
nonadiabatic generalization of the Berry phase, the 
Aharonov-Anandan phase \cite{AA87}. Our numerical results imply that
to observe clear Berry phase effects 
such as the 'magic angles' found by Engel \emph{et al.} \cite{EL00} in the magneto
conductance and UCF's of 
diffusive rings one has to go to the truly adiabatic regime given by condition 
(\ref{AC}).

For a typical experimental AB set\-up based on copper rings 
with radius $r_0=500$ nm and $\ell \!=\!15$ nm \cite{PhDH} strict
 application of the  criterion (\ref{AC}) corresponds to
$B$-field strengths larger than $10^3$ T.  On the other hand, 
according to the condition $Q\ll 1/ \sqrt{\langle N\rangle}$, 
the opposite non-adiabatic regime 
($\langle T_{\downarrow \downarrow}\rangle\! \sim\! 1$)
is restricted to fields smaller than 0.1 T. In the 
 broad intermediate $B$-field range, covering four orders of magnitude, the
magneto conductance is expected to show at most 
signatures of the Aharonov-Anandan phase.

Recently, imprints of Berry's phase in
AB oscillations have been reported for holes in quasi-ballistic 2d GaAs rings with 
strong spin-orbit interaction\cite{Yau}.
A rough estimate of the system parameters 
suggests that the experimental conditions may fulfill the adiabaticity 
criterion (\ref{AC}) with $B$ replaced by an effective Rashba field strength.

To summarize we studied spin-dependent quantum transport through  
2d disordered geometries. We showed that
the relevant parameter defining the adiabatic limit both for the total 
conductance and the quantum corrections scales with 
the square root of the number of scattering events. This can be cast into 
a generalized criterion for adiabaticity   
for both, ballistic and disordered systems. 
It appears as a severe obstacle for direct experimental 
observation of Berry phases in the conductance through diffusive metal rings. 
Our numerical findings indicate that elastic scattering due 
to non-magnetic impurities
in the presence of a spatially varying magnetic field exhibits
features similar to those in systems with 
spin-flips associated with the scattering process as 
for magnetic impurities or Rashba spin orbit coupling. 

{\em Acknowledgements:} 
We would like to thank H.-A.\ Engel and D.\ Loss for helpful discussions
and acknowledge support from the Deutsche Forschungsgemeinschaft.

%
%



\begin{thebibliography}{10}
\bibitem{comment-method}
 Testing this model on a narrow wire we find quanti\-ta\-tive 
 agreement with predictions for quantum conductance corrections 
 (weak localization, universal conductance fluctuations)
and an exponential decrease of 
 the conductance when approaching the localized regime.
\bibitem{comment-spinup}
Equivalent results are obtained for spin-up electrons.
\bibitem{comment1d} The restriction to a 1d model is justified since we 
are dealing with narrow strips of width $W \lesssim \ell$.
\bibitem{commentmodel}
To gain further insight into the scaling law (\ref{AC}) we developed 
a phenomenological model
based on a perturbative approach valid in the far adiabatic 
and non-adiabatic limit. Assuming  that
spin flips before and after a scattering event are independent processes  we
can derive a scaling behavior which coincides with condition  (\ref{AC})  \cite{PFRunp}.
The ballistic limit then corresponds to $N = 1$.
\bibitem{comment-ring} This has been numerically confirmed for the magneto conductance in
 disordered ($L / \ell \approx 4$) rings subject to an inhomogeneous field 
\cite{PFRunp}.
\end{thebibliography}
\end{document}